\documentclass[aps,pra,amsmath,amssymb,superscriptaddress,preprint]{revtex4}

\usepackage{graphicx}

\newcommand{\unit}[1]{\ensuremath{\,\mathrm{#1}}}
\newcommand{\Og}{\ensuremath{\Omega}}
\newcommand{\Om}{\ensuremath{\Omega_\mathrm{m}}}
\newcommand{\Gm}{\ensuremath{\Gamma_\mathrm{m}}}

\newcommand{\og}{\ensuremath{\omega}}
\newcommand{\meff}{m_\text{eff}}

\begin{document}


\title{Resolved-sideband cooling and measurement of a micromechanical oscillator close to the quantum limit}


\author{A.\ Schliesser}
\altaffiliation{These authors contributed equally to this work.}
\affiliation{Max Planck Institut f\"ur Quantenoptik, 85748 Garching,
Germany}
\author{O. Arcizet}
\altaffiliation{These authors contributed equally to this work.}
\affiliation{Max Planck Institut f\"ur Quantenoptik, 85748 Garching,
Germany}
\author{R.\ Rivi\`ere}
\altaffiliation{These authors contributed equally to this work.}
\affiliation{Max Planck Institut f\"ur Quantenoptik, 85748 Garching,
Germany}
\author{T.J.\ Kippenberg}
\email[]{tobias.kippenberg@epfl.ch} \affiliation{Max Planck Institut
f\"ur Quantenoptik, 85748 Garching, Germany} \affiliation{\'Ecole
Polytechnique f\'ed\'erale de Lausanne (EPFL), CH-1015 Lausanne,
Switzerland}

\date{\today}

\begin{abstract}
The observation of quantum phenomena in macroscopic mechanical
oscillators \cite{Braginsky1992, Schwab2005} has been a subject of
interest since the inception of quantum mechanics. It may provide
insights into the quantum-classical boundary, experimental
investigation of the theory of quantum measurements
\cite{Braginsky1992, Bose1999, Tittonen1999}, the origin of
mechanical decoherence \cite{Marshall2003} and generation of
non-classical states of motion. Prerequisite to this regime are both
preparation of the mechanical oscillator at low phonon occupancy and
a measurement sensitivity at the scale of the spread $\Delta x$ of
the oscillator's ground state wavefunction. Over the past decade, it
has been widely perceived that the most promising approach to
address these two challenges are electro nanomechanical systems
\cite{Schwab2005, Cleland1998, Naik2006, LaHaye2004, Regal2008,
Teufel2008}, which can be cooled with milli-Kelvin scale dilution
refrigerators, and feature large $\Delta x\sim 10^{-14}\unit{m}$
resolvable with electronic transducers such as a superconducting
single-electron transistor \cite{Naik2006,LaHaye2004,Knobel2003}, a
microwave stripline cavity \cite{Regal2008} or a quantum
interference device \cite{Etaki2008}. In this manner, thermal
occupation as low as 25 quanta \cite{Naik2006, Teufel2008} has been
measured. Here we approach for the first time the quantum regime
with a mechanical oscillator of mesoscopic dimensions--discernible
to the bare eye--and 1000-times more massive than the heaviest
nano-mechanical oscillators used to date. Imperative to these
advances are two key principles of cavity optomechanics
\cite{Kippenberg2008}: Optical interferometric measurement of
mechanical displacement at the attometer level
\cite{Arcizet2006,Schliesser2008b}, and the ability to use
measurement induced dynamic back-action
\cite{Dykman1978,Gigan2006,Arcizet2006a,Schliesser2006} to achieve
resolved sideband laser cooling \cite{Schliesser2008,Regal2008} of
the mechanical degree of freedom. Using only modest cryogenic
pre-cooling to $1.65\unit{K}$, preparation of a mechanical
oscillator close to its quantum ground state ($63\pm20$ phonons) is
demonstrated. Simultaneously, a readout sensitivity that is within a
factor of $5.5\pm1.5$ of the standard quantum limit
\cite{Braginsky1992, Caves1981} is achieved. Taking measurement
backaction into account, this represents the closest approach to the
Heisenberg uncertainty relation for continuous position measurements
yet demonstrated. The reported experiments mark a paradigm shift in
the approach to the quantum limit of mechanical oscillators using
optical techniques and represent a first step into a new era of
experimental investigation which probes the quantum nature of the
most tangible harmonic oscillator: a mechanical vibration.
\end{abstract}


\maketitle


The experimental setting of the present work is a cavity
optomechanical system, which parametrically couples optical and
mechanical degrees of freedom via radiation pressure. In the present
case, toroidal microresonators are employed which exhibit (cf. Fig.
1) strong, inherent opto-mechanical coupling between high-quality
factor ($Q>10^8$) optical whispering gallery modes and the
mechanical radial breathing mode \cite{Kippenberg2005} (RBM),
featuring high frequency (65 and 122 MHz for the resonators used in
this work), and effective masses \cite{Pinard1999} on the order of
1-10 ng (cf. Fig. 1b). The quality factors of the RBM can reach
values up to 80,000 if clamping losses are mitigated by modal
engineering \cite{Anetsberger2008}. To achieve a regime of low
mechanical oscillator occupancy we apply laser cooling to a
cryogenically pre-cooled micromechanical oscillator with high
frequency. Figure 1 shows a schematic of the experiment. A chip with
micro-resonators is inserted into a Helium exchange gas cryostat.
Piezoelectric actuators enable positioning of a tapered optical
fiber used for evanescent coupling with a resolution sufficient to
adjust the taper-toroid gap to critical coupling. The total optical
loss through the cryogenic environment can reach values below
$25\%$. Low pressure (0.1-50 mbar) Helium exchange gas is admitted
into the sample chamber, thermalizing the sample with a heat
exchanger through which ${}^4$He is pumped from a reservoir of
liquid ${}^4$He. An exchange gas temperature of 1.65 K is achieved.
Due to the low heat conductivity of  glass, and possible light
absorption, it is of prime importance to  verify the thermalization
of the mechanical oscillator to 1.65 K. To this end, we perform
noise thermometry using the RBM. A low power ($<2 \mu$W) laser is
tuned into resonance with a high-$Q$ optical mode. Fluctuations of
the cavity radius--as induced by thermal excitation of the
RBM--induce resonance frequency fluctuations of the cavity, which
are imprinted as phase fluctuations on the laser light coupling back
to the tapered fiber (cf.\ Figure 1c). A phase-sensitive detection
scheme enables measurement of the Lorentzian displacement noise
spectrum of the thermal (Brownian) motion of the RBM, characterized
by its resonance frequency $\Om$, mechanical damping rate $\Gm$  and
peak displacement amplitude $S_{xx}^{\mathrm{th}}(\Om)$  (cf.
Methods summary). Figure 2a shows the resulting mechanical mode
temperature as derived via the equipartition theorem from the
independently calibrated noise spectra, where phase-sensitive
detection was accomplished using the Pound-Drever-Hall technique.
Importantly, the temperature of the sample follows the exchange gas,
demonstrating that excellent thermalization is achieved, a key
prerequisite for the experiments described from here on. For the 62
MHz sample thermalization to 1.65 K entails an initial average
occupancy of $\langle n\rangle =k_\mathrm{B} T_\mathrm{RBM}/\hbar
\Om\approx 560$, while for the 122 MHz sample a low occupancy of
$\langle n \rangle \approx 280$ is attained. Note that despite the
modest pre-cooling to 1.65 K these occupancies are identical to
those of a 1 MHz nano-mechanical oscillator thermalized to a
dilution refrigerator temperature below 20 mK, emphasizing the
significant advantage of working with high frequency oscillators.

Measuring the mechanical displacement associated with such a massive
oscillator at low occupancies requires high sensitivity, in
particular, since the mechanical quality factor of silica is reduced
to $\sim2000$ at 1.65 K due to losses originating from phonon
coupling to structural defect states \cite{Arcizet2009a} (note that
damping by the exchange gas is negligible, and that the mechanical Q
factor improves again at lower temperatures \cite{Pohl2002}). The
required attometer-level sensitivity can (so far) only be achieved
with optical transducers. Following our previous work, we employ
homodyne spectroscopy \cite{Schliesser2008b} based on a
quantum-noise limited titanium sapphire laser (in both amplitude and
phase), which is resonantly coupled to WGM resonance in the vicinity
of a wavelength of 780 nm. The laser's phase shift introduced by the
mechanical fluctuations are detected interferometrically, by
comparison with a high-power (2-5 mW) optical phase reference (local
oscillator). Frequency analysis of this signal yields the thermal
noise displacement spectrum $S_{xx}^{\mathrm{th}}(\Og)$, on top of a
measurement background. Figure 2b shows data obtained from the RBM
of a $55 \mu\mathrm{m}$-diameter micro-resonator cooled to T=2.4 K,
or $\langle n \rangle \approx 770$. The background of this
measurement is at a level of $~1.5\cdot
10^{-18}\unit{m}/\sqrt{\mathrm{Hz}}$, which is only a factor of
$5.5\pm 1.5$-times higher than the standard quantum limit (SQL)
\cite{Braginsky1992, Caves1981}, given by
$\sqrt{S_{xx}^{\mathrm{SQL}}(\Om)}=\sqrt{\hbar/\meff \Gm \Om}$ for a
measurement at the mechanical resonance. This proves the
counterintuitive notion, that measurements close to the standard
quantum limit are possible, in spite of the here used strategy of
using high frequency and comparatively massive oscillators  (in
contrast to nanomechanical systems).

In order to further decrease the number of thermal quanta of the
mechanical oscillator we use cooling via radiation pressure
dynamical backaction as predicted \cite{Dykman1978,Braginsky2002}
and recently experimentally demonstrated \cite{Gigan2006,
Arcizet2006a, Schliesser2006}. Similar to the atomic physics
\cite{Wineland1979} case, ground state cooling requires accessing
the resolved sideband regime \cite{Schliesser2008,
Wilson-Rae2007,Marquardt2007}, which necessitates the mechanical
oscillator frequency to exceed the cavity decay rate (i.e.\
$\Om\gg\kappa$). This regime is moreover prerequisite for schemes
such as two transducer quantum non-demolition (QND) measurements
\cite{Braginsky1992, Braginsky1996} or the preparation of a
mechanical oscillator in a squeezed state of motion
\cite{Clerk2008a}. Operation in the RSB regime as demonstrated in
\cite{Schliesser2008, Teufel2008} is accomplished using a cavity
with a narrow resonance (5.5~MHz intrinsic decay rate and 9~MHz mode
splitting), which is broadened to a $\kappa/2\pi\approx
19\unit{MHz}$-wide resonance due to fiber coupling (corresponding to
a loaded finesse of $\sim 70,000$). The laser is subsequently tuned
to the lower mechanical sideband, i.e.\ red-detuned by 65.2 MHz, the
resonance frequency of this sample's RBM. For this detuning the
circulating power is reduced by a factor of $4\Om^2/\kappa^2$. At
the same time, the sensitivity to mechanical displacements is
slightly reduced. In the ideal case of a highly overcoupled cavity,
with unity detection efficiency and no excess noise except for the
laser's intrinsic quantum noise, the imprecision noise spectral
density, i.e. the background of the measurement caused by detection
shot noise, is given by:
\begin{equation}
  S_{xx}(\Og)=\frac{\hbar \og}{16 g_0^2 \cdot P}
  \left(\frac{\Delta^2+(\kappa/2)^2}{\kappa/2}\right)^2
  \left(1+\Og^2 \frac{\Og^2+(\kappa/2)^2-2 \Delta^2}
    {(\Delta^2+(\kappa/2)^2)^2+\Og^2(\kappa/2)^2}\right)
\end{equation} where the opto-mechanical coupling $g_0=d\og/dx$,
and $d\og/dx=\og/R$  in the present embodiment. Moreover, $R$ is the
cavity radius, $P$ the launched input laser power, $\og/2\pi$ the
optical resonance frequency, $\Delta/2 \pi$ the detuning from the
cavity resonance, and $\Og/2\pi$ the analysis frequency. In the
resolved sideband case $\Om\gg\kappa$, this expression simplifies to
$S_{xx}(\Om)=\hbar \og \Om^2/4 g_0^2 P$ at the mechanical resonance
frequency, when detuned to the first (upper or lower) mechanical
sideband $|\Delta|=\Om$. This is only a factor of 4 higher than in
the resonant readout case $\Delta=0$. We note the interesting result
that this expression does not depend on optical finesse in the
deeply resolved sideband regime.

As shown in previous work \cite{Teufel2008, Schliesser2008}, the
laser detuned to the lower sideband leads to a significant reduction
of the thermal occupation (i.\ e.\ cooling), as evidenced by the
reduced area underneath the peaks associated with the oscillator's
thermal noise (cf. Fig. 3b). The underlying physical mechanism
giving rise to cooling is enhanced anti-Stokes scattering into the
cavity mode, whereby each scattering process annihilates a thermal
phonon. In the resolved-sideband regime, ground state cooling is
possible in principle, and the minimum occupation that can be
reached is given by $\langle \tilde n
\rangle\cong\kappa^2/16\Om^2\ll1$ \cite{Wilson-Rae2007,
Marquardt2007}. As the laser cools the resonator out of equilibrium
with the thermal bath (at temperature $T$), however, heating through
the bath competes with laser cooling and leads to a final occupation
of $\langle n_\mathrm{f}\rangle\approx
\frac{\Gm}{\Gm+\Gamma_\mathrm{cool}}k_\mathrm{B} T/\hbar \Om$, where
$\Gm$ is the intrinsic damping rate and $\Gamma_\mathrm{cool}$ the
laser induced cooling rate. In the case of the data shown in Fig. 3,
a strong increase in the damping with a concomitant reduction of the
thermal occupation can be observed. The highest attained total
damping rate is  $(\Gm+\Gamma_\mathrm{cool})/2 \pi=370\unit{kHz}$,
reached with a launched power of $\sim 0.2\unit{mW}$. Evaluation of
the calibrated thermal noise spectrum reveals an effective mode
temperature of $200\pm60 \unit{mK}$, which corresponds to an average
occupation as low as $\langle n_\mathrm{f}\rangle=63\pm20$ quanta.
This is the lowest reported occupancy for a cavity optomechanical
cooling experiment reported to date; slightly lower occupancy,
$\langle n_\mathrm{f}\rangle=25$, has only been attained in the
context of conventional dilution refrigeration of nanomechanical
oscillators \cite{Naik2006}, albeit with a signal-to-background
ratio well below unity. Back-action cooling techniques applied to
nano-mechanical oscillators at milli-Kelvin temperatures achieve
occupation numbers about one order of magnitude higher
\cite{Naik2006, Teufel2008}.

Moreover, a key aspect of the reported experiments is operation in
the resolved sideband regime. To illustrate the instrumental role of
the RSB regime for accessing low phonon occupation number of
mechanical oscillators, we compare the cooling run just described
with a further, independent run with a smaller sample ($\Om/2\pi=122
\unit{MHz}$), exhibiting significantly broader linewidth (fiber
coupling broadened to $\kappa/2\pi=155\unit{MHz}$), as shown in Fig.
4. In this case, a significant deviation from the linear cooling
behavior is observed, with higher temperatures measured than
expected based on the cooling rate. It is possible to model this
deviation by taking an intracavity-power dependent heating of the
mechanical mode into account. Heating of the mechanical mode can be
caused by both quantum \cite{Braginsky1992, Genes2008} or classical
fluctuations \cite{Schliesser2008} of the radiation pressure force.
While the latter is ruled out in our experiment by the use of a
low-noise laser source (quantum limited in amplitude and phase for
the Fourier frequency of interest), the former effect cannot account
for the heating due to its much smaller magnitude, as detailed
below. Instead, the effect is attributed to heating via laser
absorption. Both data sets exhibit heating of ca. 10 K/Watt of
circulating optical power. While quantitative modeling would have to
take characteristics of the optical mode and heat transfer in the
gas-cooled sample into account, we note that similar values of laser
induced heating were extracted from studies of the optical
bistability at low temperature at a wavelength of 1.5 $\mu$m
\cite{Arcizet2009a}. Indeed, operating in the resolved-sideband
regime \cite{Schliesser2008} allows to mitigate this effect and
enables the results demonstrated in this work, as well as future
work aiming at achieving ground-state cooling, as it strongly
reduces the circulating intracavity power.

It is also interesting to consider the reported measurements from
the perspective of the theory of quantum measurement
\cite{Braginsky1992, Clerk2008}. For linear continuous measurements
as those demonstrated here, the total measurement uncertainty arises
from two intrinsic sources of noise: measurement imprecision and
measurement backaction. Measurement imprecision arises from
fluctuations at the output of the measurement device, which are not
related to mechanical oscillator motion. In the case of an optical
interferometric measurement as reported here, measurement
imprecision can be reduced to the shot noise in the detection
process (cf. eq. (1)). Measurement backaction, on the other hand,
describes the perturbation of the mechanical oscillator by the
process of the measurement. For a mechanical oscillator, this occurs
in the form of a fluctuating force, characterized by a spectral
density $S_{FF}(\Og)$. Quantum mechanics poses strict limits on how
small $S_{xx}(\Og)$  and $S_{FF}(\Og)$ can be; the product obeying
$\sqrt{S_{xx}(\Og)\cdot S_{FF}(\Og)}\geq \hbar/2$ where an equality
can be reached in the ideal case and in the absence of correlations
of imprecision and backaction noise \cite{Braginsky1992}. While
classical fluctuations (caused by laser heating or classical laser
amplitude and phase noise) can increase the value of the measurement
backaction, the lowest possible force noise for a coherent input is
given by the quantum backaction (QBA):
\begin{equation}
S_{FF}^{\mathrm{qba}}(\Og)=\frac{\hbar}{2\og}g_0^2 P_\mathrm{in}
\left(\frac{\kappa^2}{(\kappa/2)^2+\Delta^2}\right)
\left(\frac{1}{(\kappa/2)^2+(\Delta-\Om)^2}+\frac{1}{(\kappa/2)^2+(\Delta+\Om)^2}\right)
\end{equation}
where $S_{FF}^{\mathrm{qba}}(\Om)=2 g_0^2 P_\mathrm{in} \hbar/\og
\Om^2$ in the deeply resolved sideband regime
($|\Delta|=\Om\gg\kappa$). For the parameters of the measurement
with $\langle n_\mathrm{f}\rangle=63\pm20$, we find
$\sqrt{S_{FF}^{\mathrm{qba}}(\Om)}\approx 1
\unit{fN}/\sqrt{\mathrm{Hz}}$. For the present experiments, the
dominant measurement backaction effect is the heating by laser
absorption (as observed in Fig. 4), which increases the magnitude of
the thermal Langevin force. Indeed, using the effective mechanical
susceptibility (at resonance) of the laser-cooled oscillator
$|\chi_\mathrm{eff}(\Om)|=(\meff(\Gm+\Gamma_\mathrm{cool})\Om)^{-1}$,
we can retrieve the magnitude of the \emph{total} force spectral
density from the measurement of its displacement spectrum via
$S_{FF}^{\mathrm{tot}}(\Om)=S_{xx}^{\mathrm{th}}(\Om)/|\chi_\mathrm{eff}(\Om)|^2$.
It amounts to $\sqrt{S_{FF}^{\mathrm{tot}}(\Om)}\approx 8
\unit{fN}/\sqrt{\mathrm{Hz}}$ and is dominated by the thermal
Langevin force associated with the toroid's temperature. While only
a part of this force is caused by the measurement process via the
heating induced temperature increase, we can nonetheless calculate
the product $\sqrt{S_{xx}(\Om)\cdot S_{FF}^{\mathrm{tot}}(\Om)}$ as
an upper bound, of how closely the ideal case has been approached in
these measurements. For this the conservative assumption that the
entire force spectral density acting on the mechanical oscillator
(including the thermal Langevin force) is caused by measurement
backaction is used. Using $\sqrt{S_{xx}(\Og)}\approx
1.4\cdot10^{-18} \unit{m}/\sqrt{\mathrm{Hz}}$ we thereby find a
product of $\sqrt{S_{xx}(\Om)\cdot S_{FF}^{\mathrm{tot}}(\Om)}$ as
low as $230 \cdot \hbar/2$. This represents a factor of 4
improvement compared to the closest approach made with a
nano-electromechanical system \cite{Naik2006,Clerk2008}. We expect
that improvements on our results are readily attainable by operating
more deeply in the resolved-sideband regime \cite{Schliesser2008},
in a colder cryogenic environment (such as ${}^3\mathrm{He}$), and
with reduced mechanical dissipation by employing e.g. crystalline
resonators \cite{Ilchenko2004}. The properties of the system
demonstrated here--a massive oscillator, prepared with very low
average occupation, which is coupled to an ultra low loss fiber
transport medium is pivotal for a variety of experiments and
protocols involving photons and phonons, such as radiation pressure
squeezing \cite{Fabre1994, Mancini1994}, entanglement
\cite{Vitali2007} or QND measurements \cite{Clerk2008a,
Heidmann1997}.

In summary, our result signal a paradigm shift in the access to the
quantum regime of mechanical oscillators by demonstrating the clear
feasibility to enter this regime using significantly more massive
cavity optomechanical \cite{Kippenberg2008, Kippenberg2007} systems
and using modest pre-cooling; a regime so far only attained with
nanomechanical oscillator thermalized to milli-Kelvin scale dilution
refridgerator temperatures. As such, the reported experiments-which
demonstrate preparation and sensitive readout of mechanical
oscillator with few quanta-mark a first step into a new era of
experimental investigation of mechanical systems in the quantum
regime, which has applications ranging from fundamental predictions
of quantum measurement theory tests of quantum mechanics to the
generation of non-classical states of motion and study of mechanical
decoherence.

 \begin{figure}[tbh]
\includegraphics[width=.8\linewidth]{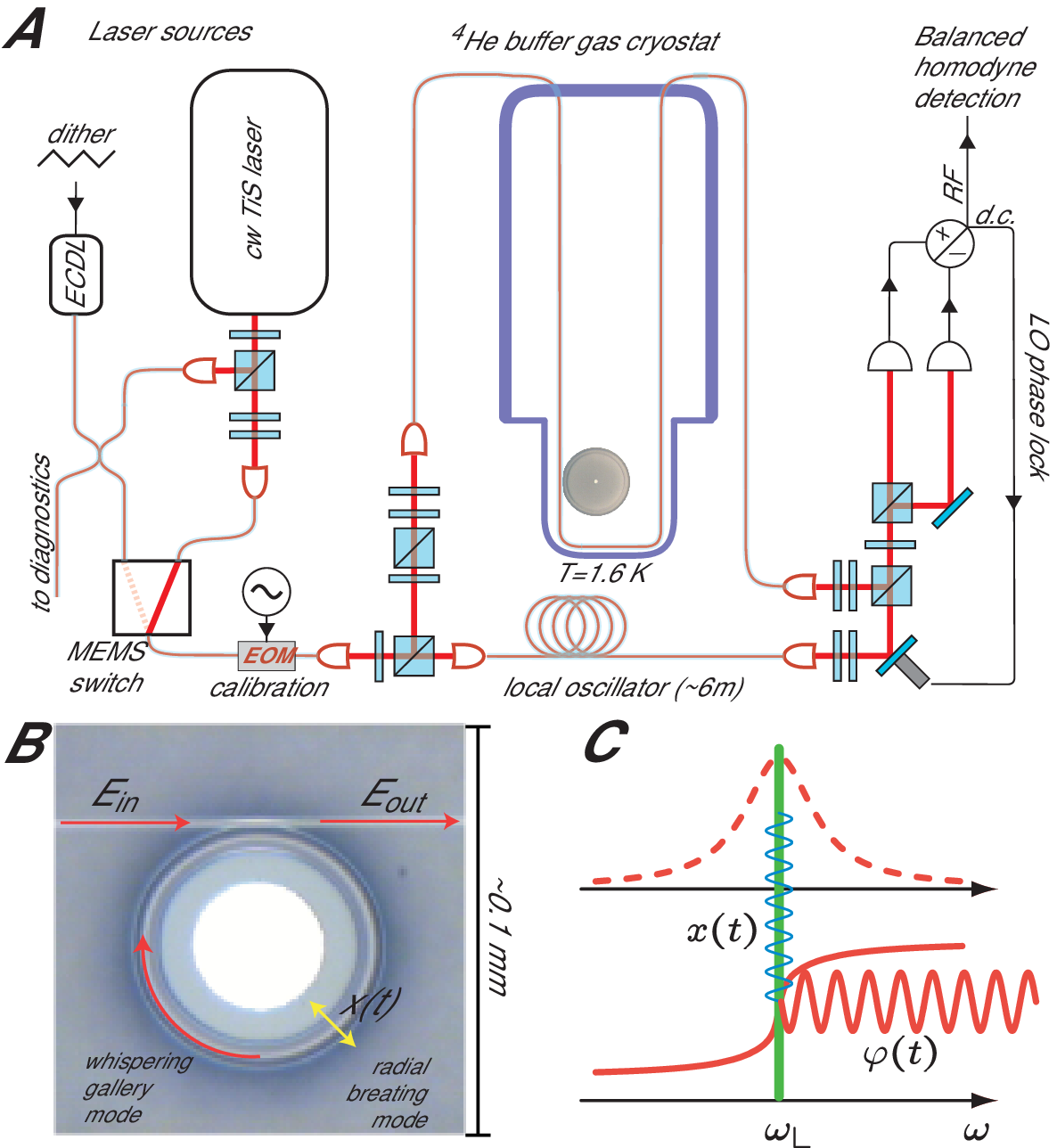}%
 \caption{\label{f:setup}%
 Cryogenic cooling and displacement measurements of a micromechanical oscillator.
 (A to C) A silica microtoroid is held in a 1.65 K-cold ${}^4$He atmosphere. The
 toroid supports both high-Q optical whispering gallery modes and a mechanical
 radial breathing mode, which are parametrically coupled to an optical resonance
 frequency shift induced by mechanical displacement. High-Q optical resonances are
 identified using a tunable external-cavity diode laser (ECDL). To probe the
  mechanical oscillator, the light input is switched to a low-noise Ti:Sapphire. A
   small fraction of the laser beam is sent into the cryostat and couples to the
   WGM by evanescent coupling from a fiber taper approached to the rim of the toroid.
    Balanced homodyne measurement of the laser phase as transmitted through the taper
     is implemented using a Mach-Zehnder fiber interferometer (phase plates and
     polarizing beam splitters are only schematically indicated). A modulation
     $x(t)$ of the radius of the cavity (C) induces a modulation of the phase of the light $\phi(t)$
      emerging from the cavity. This phase shift is detected by comparison with a phase reference, derived from the same laser in a beam splitter followed by a balanced detector.}
 \end{figure}

\begin{figure}[tbh]
 \includegraphics[width=\linewidth]{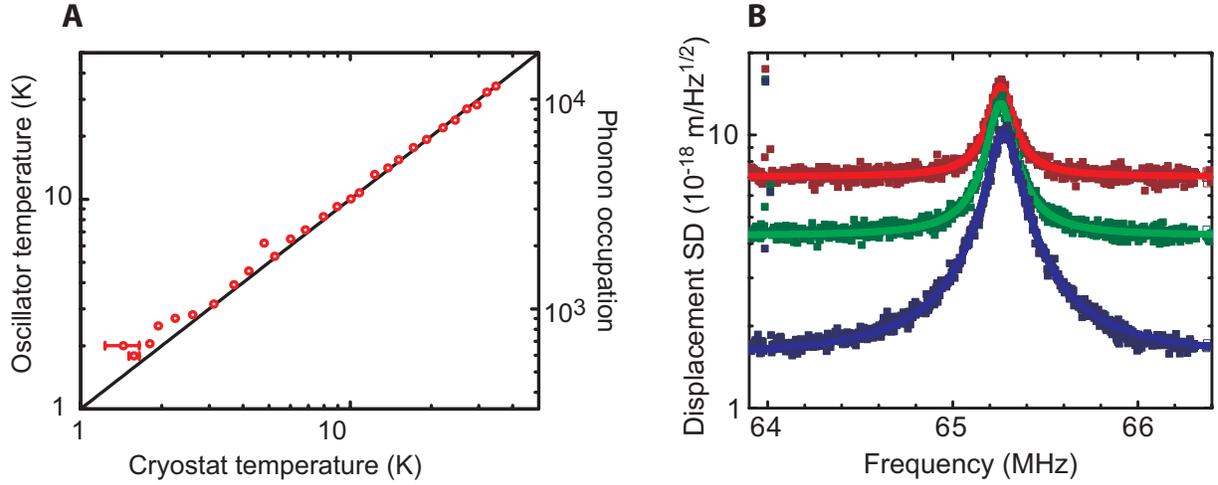}%
 \caption{\label{f:setup}%
 Thermalization and probing of a micromechanical oscillator (A) Using noise thermometry,
 the temperature of the mechanical oscillator is determined as a function of the
 temperature of the cryostat (exchange gas). The noise temperature of the 62 MHz mode
  follows the cryostat temperature in a linear manner, down to an occupation of less than 1000 phonons. (B)
  In spite of the low phonon occupation (770 in the case of this 65.3 MHz oscillator) displacement
   monitoring with high signal-to-noise ratio is possible using optical techniques (at the attometer
    level in the present case). The optical power used to interrogate the mechanical oscillator was
     ca.\ $3 \unit{\mu W}$ (red trace), $10 \unit{\mu W}$ (green trace) and $100 \unit{\mu W}$ (blue trace). The measurement background is given by shot noise in the detection.}
\end{figure}

  \begin{figure}[tbh]
 \includegraphics[width=\linewidth]{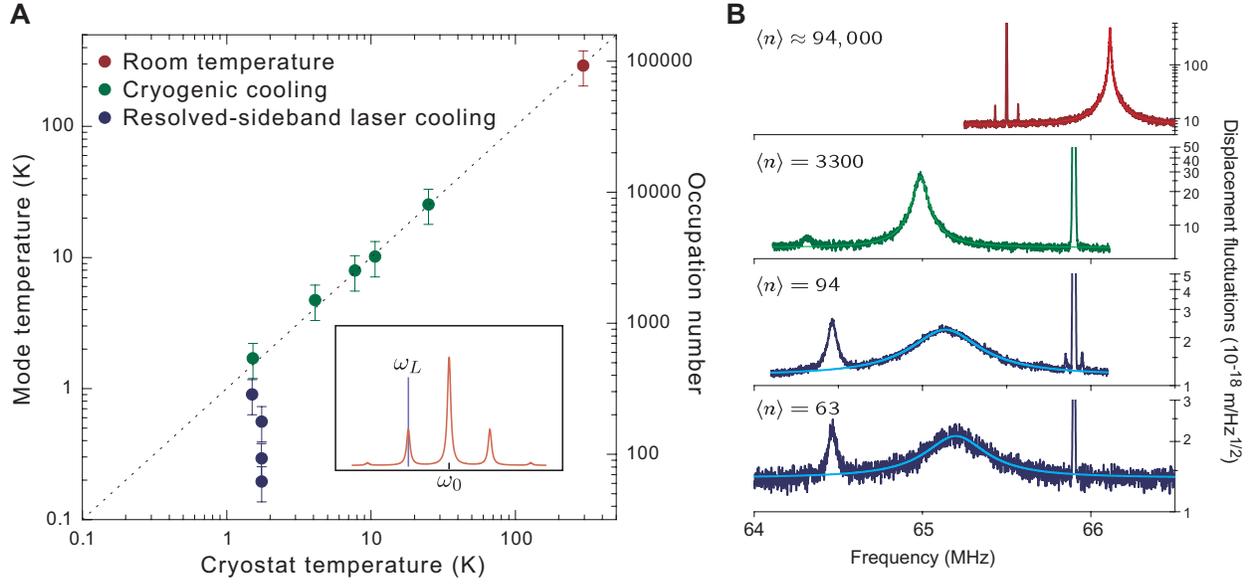}%
 \caption{\label{f:setup}
 Cryogenic precooling and resolved-sideband laser cooling. (A) Combined with pre-cooling in the cryostat to $1.65\unit{K}$,
  resolved-sideband cooling reduces the occupation of the oscillator further to
  $\langle n_\mathrm{f} \rangle=63\pm 20$ phonons. The inset illustrates the laser (blue line) detuned to the lower mechanical sideband of the cavity spectrum (red line). (B) Displacement noise spectra of the RBM with four different average occupation numbers, together with Lorentzian fit, at room temperature (red curve) after cryogenic cooling to 10K (green curve), and after additional resolved sideband cooling (blue curves). The sharp calibration peak, and a second mode at slightly lower frequency are also shown. }
 \end{figure}

  \begin{figure}[tbh]
\includegraphics[width=\linewidth]{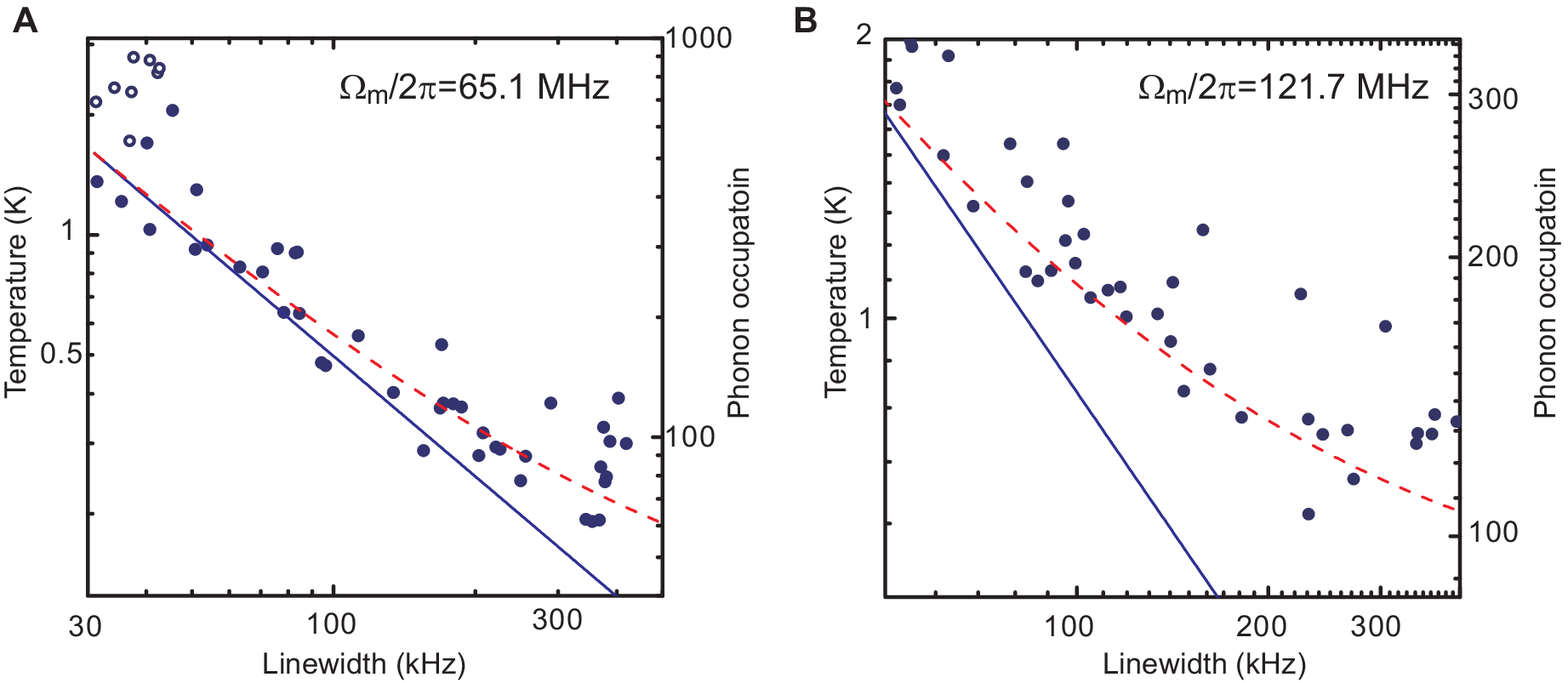}%
 \caption{\label{f:setup}%
Resolved sideband cooling of (A) a RBM at 65.1 MHz and (B) an RBM at
121.7 MHz frequency. The mode temperature and corresponding phonon
occupation are reduced as the detuned laser induces additional
damping, and therefore increases the linewidth of the thermal noise
spectrum. Full points correspond to mechanical spectra taken with a
detuned cooling laser, open points in (A) correspond to measurements
with the laser tuned close to the optical resonance. The deviation
from a linear cooling behaviour (blue line) indicates a backaction
heating effect, which is compatible with heating of the structure
originating from residual absorption (red dashed line). This effect
is suppressed in the resolved sideband regime (panel (A)).}
 \end{figure}

\begin{acknowledgments}
This work was supported by an Independent Max Planck Junior Research
Group of the Max Planck Society, the Deutsche Forschungsgemeinschaft
(DFG-GSC) and a Marie Curie Excellence Grant, and the FP7 project
"MINOS". O.\ A.\ acknowledges funding from a Marie Curie Grant
(project QUOM). Thomas Becker is gratefully acknowledged for support
with the cryogenic experiments, and J\"org Kotthaus for sample
fabrication.
\end{acknowledgments}

\bibliographystyle{apsrev}
\bibliography{C:/Dokume\string~1/Albert/Eigene\string~1/Literature/microcavities}

\begin{thebibliography}{40}
\expandafter\ifx\csname natexlab\endcsname\relax\def\natexlab#1{#1}\fi
\expandafter\ifx\csname bibnamefont\endcsname\relax
  \def\bibnamefont#1{#1}\fi
\expandafter\ifx\csname bibfnamefont\endcsname\relax
  \def\bibfnamefont#1{#1}\fi
\expandafter\ifx\csname citenamefont\endcsname\relax
  \def\citenamefont#1{#1}\fi
\expandafter\ifx\csname url\endcsname\relax
  \def\url#1{\texttt{#1}}\fi
\expandafter\ifx\csname urlprefix\endcsname\relax\def\urlprefix{URL }\fi
\providecommand{\bibinfo}[2]{#2}
\providecommand{\eprint}[2][]{\url{#2}}

\bibitem[{\citenamefont{Braginsky and Khalili}(1992)}]{Braginsky1992}
\bibinfo{author}{\bibfnamefont{V.~B.} \bibnamefont{Braginsky}}
  \bibnamefont{and} \bibinfo{author}{\bibfnamefont{F.~Y.}
  \bibnamefont{Khalili}}, \emph{\bibinfo{title}{Quantum {M}easurement}}
  (\bibinfo{publisher}{Cambridge University Press}, \bibinfo{year}{1992}).

\bibitem[{\citenamefont{Schwab and Roukes}(2005)}]{Schwab2005}
\bibinfo{author}{\bibfnamefont{K.~C.} \bibnamefont{Schwab}} \bibnamefont{and}
  \bibinfo{author}{\bibfnamefont{M.~L.} \bibnamefont{Roukes}},
  \bibinfo{journal}{Physics Today} \textbf{\bibinfo{volume}{58}},
  \bibinfo{pages}{36} (\bibinfo{year}{2005}).

\bibitem[{\citenamefont{Bose et~al.}(1999)\citenamefont{Bose, Jacobs, and
  Knight}}]{Bose1999}
\bibinfo{author}{\bibfnamefont{S.}~\bibnamefont{Bose}},
  \bibinfo{author}{\bibfnamefont{K.}~\bibnamefont{Jacobs}}, \bibnamefont{and}
  \bibinfo{author}{\bibfnamefont{P.~L.} \bibnamefont{Knight}},
  \bibinfo{journal}{Physical Review A} \textbf{\bibinfo{volume}{59}},
  \bibinfo{pages}{3204} (\bibinfo{year}{1999}).

\bibitem[{\citenamefont{Tittonen et~al.}(1999)\citenamefont{Tittonen,
  Breitenbach, Kalkbrenner, M{\"u}ller, Conradt, Schiller, Steinsland, Blanc,
  and de~Rooij}}]{Tittonen1999}
\bibinfo{author}{\bibfnamefont{I.}~\bibnamefont{Tittonen}},
  \bibinfo{author}{\bibfnamefont{G.}~\bibnamefont{Breitenbach}},
  \bibinfo{author}{\bibfnamefont{T.}~\bibnamefont{Kalkbrenner}},
  \bibinfo{author}{\bibfnamefont{T.}~\bibnamefont{M{\"u}ller}},
  \bibinfo{author}{\bibfnamefont{R.}~\bibnamefont{Conradt}},
  \bibinfo{author}{\bibfnamefont{S.}~\bibnamefont{Schiller}},
  \bibinfo{author}{\bibfnamefont{E.}~\bibnamefont{Steinsland}},
  \bibinfo{author}{\bibfnamefont{N.}~\bibnamefont{Blanc}}, \bibnamefont{and}
  \bibinfo{author}{\bibfnamefont{N.~F.} \bibnamefont{de~Rooij}},
  \bibinfo{journal}{Physical Review A} \textbf{\bibinfo{volume}{59}},
  \bibinfo{pages}{1038} (\bibinfo{year}{1999}).

\bibitem[{\citenamefont{Marshall et~al.}(2003)\citenamefont{Marshall, Simon,
  Penrose, and Bouwmeester}}]{Marshall2003}
\bibinfo{author}{\bibfnamefont{W.}~\bibnamefont{Marshall}},
  \bibinfo{author}{\bibfnamefont{C.}~\bibnamefont{Simon}},
  \bibinfo{author}{\bibfnamefont{R.}~\bibnamefont{Penrose}}, \bibnamefont{and}
  \bibinfo{author}{\bibfnamefont{D.}~\bibnamefont{Bouwmeester}},
  \bibinfo{journal}{Physical Review Letters} \textbf{\bibinfo{volume}{91}},
  \bibinfo{pages}{130401} (\bibinfo{year}{2003}).

\bibitem[{\citenamefont{Cleland and Roukes}(1998)}]{Cleland1998}
\bibinfo{author}{\bibfnamefont{A.}~\bibnamefont{Cleland}} \bibnamefont{and}
  \bibinfo{author}{\bibfnamefont{M.}~\bibnamefont{Roukes}},
  \bibinfo{journal}{Nature} \textbf{\bibinfo{volume}{392}},
  \bibinfo{pages}{160} (\bibinfo{year}{1998}).

\bibitem[{\citenamefont{Naik et~al.}(2006)\citenamefont{Naik, Buu, LaHaye,
  Armour, Clerk, Blencowe, and Schwab}}]{Naik2006}
\bibinfo{author}{\bibfnamefont{A.}~\bibnamefont{Naik}},
  \bibinfo{author}{\bibfnamefont{O.}~\bibnamefont{Buu}},
  \bibinfo{author}{\bibfnamefont{M.~D.} \bibnamefont{LaHaye}},
  \bibinfo{author}{\bibfnamefont{A.~D.} \bibnamefont{Armour}},
  \bibinfo{author}{\bibfnamefont{A.~A.} \bibnamefont{Clerk}},
  \bibinfo{author}{\bibfnamefont{M.~P.} \bibnamefont{Blencowe}},
  \bibnamefont{and} \bibinfo{author}{\bibfnamefont{K.~C.}
  \bibnamefont{Schwab}}, \bibinfo{journal}{Nature}
  \textbf{\bibinfo{volume}{443}}, \bibinfo{pages}{193} (\bibinfo{year}{2006}).

\bibitem[{\citenamefont{LaHaye et~al.}(2004)\citenamefont{LaHaye, Buu,
  Camarota, and Schwab}}]{LaHaye2004}
\bibinfo{author}{\bibfnamefont{M.~D.} \bibnamefont{LaHaye}},
  \bibinfo{author}{\bibfnamefont{O.}~\bibnamefont{Buu}},
  \bibinfo{author}{\bibfnamefont{B.}~\bibnamefont{Camarota}}, \bibnamefont{and}
  \bibinfo{author}{\bibfnamefont{K.~C.} \bibnamefont{Schwab}},
  \bibinfo{journal}{Science} \textbf{\bibinfo{volume}{304}},
  \bibinfo{pages}{74} (\bibinfo{year}{2004}).

\bibitem[{\citenamefont{Regal et~al.}(2008)\citenamefont{Regal, Teufel, and
  Lehnert}}]{Regal2008}
\bibinfo{author}{\bibfnamefont{C.~A.} \bibnamefont{Regal}},
  \bibinfo{author}{\bibfnamefont{J.~D.} \bibnamefont{Teufel}},
  \bibnamefont{and} \bibinfo{author}{\bibfnamefont{K.~W.}
  \bibnamefont{Lehnert}}, \bibinfo{journal}{Nature Physics}
  \textbf{\bibinfo{volume}{4}}, \bibinfo{pages}{555} (\bibinfo{year}{2008}).

\bibitem[{\citenamefont{Teufel et~al.}(2008)\citenamefont{Teufel, Harlow,
  Regal, and Lehnert}}]{Teufel2008}
\bibinfo{author}{\bibfnamefont{J.~D.} \bibnamefont{Teufel}},
  \bibinfo{author}{\bibfnamefont{J.~D.} \bibnamefont{Harlow}},
  \bibinfo{author}{\bibfnamefont{C.~A.} \bibnamefont{Regal}}, \bibnamefont{and}
  \bibinfo{author}{\bibfnamefont{K.~W.} \bibnamefont{Lehnert}},
  \bibinfo{journal}{Physical Review Letters} \textbf{\bibinfo{volume}{101}},
  \bibinfo{pages}{197203} (\bibinfo{year}{2008}).

\bibitem[{\citenamefont{Knobel and Cleland}(2003)}]{Knobel2003}
\bibinfo{author}{\bibfnamefont{R.~G.} \bibnamefont{Knobel}} \bibnamefont{and}
  \bibinfo{author}{\bibfnamefont{A.~N.} \bibnamefont{Cleland}},
  \bibinfo{journal}{Nature} \textbf{\bibinfo{volume}{424}},
  \bibinfo{pages}{291} (\bibinfo{year}{2003}).

\bibitem[{\citenamefont{Etaki et~al.}(2008)\citenamefont{Etaki, Poot, Mahbobb,
  Onomitsu, Yamaguchi, and van~der Zant}}]{Etaki2008}
\bibinfo{author}{\bibfnamefont{S.}~\bibnamefont{Etaki}},
  \bibinfo{author}{\bibfnamefont{M.}~\bibnamefont{Poot}},
  \bibinfo{author}{\bibfnamefont{I.}~\bibnamefont{Mahbobb}},
  \bibinfo{author}{\bibfnamefont{K.}~\bibnamefont{Onomitsu}},
  \bibinfo{author}{\bibfnamefont{H.}~\bibnamefont{Yamaguchi}},
  \bibnamefont{and} \bibinfo{author}{\bibfnamefont{H.~S.~J.}
  \bibnamefont{van~der Zant}}, \bibinfo{journal}{Nature Physics}
  \textbf{\bibinfo{volume}{4}}, \bibinfo{pages}{785} (\bibinfo{year}{2008}).

\bibitem[{\citenamefont{Kippenberg and Vahala}(2008)}]{Kippenberg2008}
\bibinfo{author}{\bibfnamefont{T.~J.} \bibnamefont{Kippenberg}}
  \bibnamefont{and} \bibinfo{author}{\bibfnamefont{K.~J.}
  \bibnamefont{Vahala}}, \bibinfo{journal}{Science}
  \textbf{\bibinfo{volume}{321}}, \bibinfo{pages}{1172} (\bibinfo{year}{2008}).

\bibitem[{\citenamefont{Arcizet
  et~al.}(2006{\natexlab{a}})\citenamefont{Arcizet, Cohadon, Briant, Pinard,
  Heidmann, Mackowski, Michel, Pinard, Fran{c}ais, and Rousseau}}]{Arcizet2006}
\bibinfo{author}{\bibfnamefont{O.}~\bibnamefont{Arcizet}},
  \bibinfo{author}{\bibfnamefont{P.-F.} \bibnamefont{Cohadon}},
  \bibinfo{author}{\bibfnamefont{T.}~\bibnamefont{Briant}},
  \bibinfo{author}{\bibfnamefont{M.}~\bibnamefont{Pinard}},
  \bibinfo{author}{\bibfnamefont{A.}~\bibnamefont{Heidmann}},
  \bibinfo{author}{\bibfnamefont{J.-M.} \bibnamefont{Mackowski}},
  \bibinfo{author}{\bibfnamefont{C.}~\bibnamefont{Michel}},
  \bibinfo{author}{\bibfnamefont{L.}~\bibnamefont{Pinard}},
  \bibinfo{author}{\bibfnamefont{O.}~\bibnamefont{Fran{c}ais}},
  \bibnamefont{and} \bibinfo{author}{\bibfnamefont{L.}~\bibnamefont{Rousseau}},
  \bibinfo{journal}{Physical Review Letters} \textbf{\bibinfo{volume}{97}},
  \bibinfo{pages}{133601} (\bibinfo{year}{2006}{\natexlab{a}}).

\bibitem[{\citenamefont{Schliesser
  et~al.}(2008{\natexlab{a}})\citenamefont{Schliesser, Anetsberger,
  Rivi{\`e}re, Arcizet, and Kippenberg}}]{Schliesser2008b}
\bibinfo{author}{\bibfnamefont{A.}~\bibnamefont{Schliesser}},
  \bibinfo{author}{\bibfnamefont{G.}~\bibnamefont{Anetsberger}},
  \bibinfo{author}{\bibfnamefont{R.}~\bibnamefont{Rivi{\`e}re}},
  \bibinfo{author}{\bibfnamefont{O.}~\bibnamefont{Arcizet}}, \bibnamefont{and}
  \bibinfo{author}{\bibfnamefont{T.~J.} \bibnamefont{Kippenberg}},
  \bibinfo{journal}{New Journal of Physics} \textbf{\bibinfo{volume}{10}},
  \bibinfo{pages}{095015} (\bibinfo{year}{2008}{\natexlab{a}}).

\bibitem[{\citenamefont{Dykman}(1978)}]{Dykman1978}
\bibinfo{author}{\bibfnamefont{M.~I.} \bibnamefont{Dykman}},
  \bibinfo{journal}{Sov. Phys. Solid State} \textbf{\bibinfo{volume}{20}},
  \bibinfo{pages}{1306} (\bibinfo{year}{1978}).

\bibitem[{\citenamefont{Gigan et~al.}(2006)\citenamefont{Gigan, B{\"o}hm,
  Paternosto, Blaser, Langer, Hertzberg, Schwab, B{\"a}uerle, Aspelmeyer, and
  Zeilinger}}]{Gigan2006}
\bibinfo{author}{\bibfnamefont{S.}~\bibnamefont{Gigan}},
  \bibinfo{author}{\bibfnamefont{H.~R.} \bibnamefont{B{\"o}hm}},
  \bibinfo{author}{\bibfnamefont{M.}~\bibnamefont{Paternosto}},
  \bibinfo{author}{\bibfnamefont{F.}~\bibnamefont{Blaser}},
  \bibinfo{author}{\bibfnamefont{G.}~\bibnamefont{Langer}},
  \bibinfo{author}{\bibfnamefont{J.~B.} \bibnamefont{Hertzberg}},
  \bibinfo{author}{\bibfnamefont{K.~C.} \bibnamefont{Schwab}},
  \bibinfo{author}{\bibfnamefont{D.}~\bibnamefont{B{\"a}uerle}},
  \bibinfo{author}{\bibfnamefont{M.}~\bibnamefont{Aspelmeyer}},
  \bibnamefont{and}
  \bibinfo{author}{\bibfnamefont{A.}~\bibnamefont{Zeilinger}},
  \bibinfo{journal}{Nature} \textbf{\bibinfo{volume}{444}}, \bibinfo{pages}{67}
  (\bibinfo{year}{2006}).

\bibitem[{\citenamefont{Arcizet
  et~al.}(2006{\natexlab{b}})\citenamefont{Arcizet, Cohadon, Briant, Pinard,
  and Heidmann}}]{Arcizet2006a}
\bibinfo{author}{\bibfnamefont{O.}~\bibnamefont{Arcizet}},
  \bibinfo{author}{\bibfnamefont{P.-F.} \bibnamefont{Cohadon}},
  \bibinfo{author}{\bibfnamefont{T.}~\bibnamefont{Briant}},
  \bibinfo{author}{\bibfnamefont{M.}~\bibnamefont{Pinard}}, \bibnamefont{and}
  \bibinfo{author}{\bibfnamefont{A.}~\bibnamefont{Heidmann}},
  \bibinfo{journal}{Nature} \textbf{\bibinfo{volume}{444}}, \bibinfo{pages}{71}
  (\bibinfo{year}{2006}{\natexlab{b}}).

\bibitem[{\citenamefont{Schliesser et~al.}(2006)\citenamefont{Schliesser,
  Del'Haye, Nooshi, Vahala, and Kippenberg}}]{Schliesser2006}
\bibinfo{author}{\bibfnamefont{A.}~\bibnamefont{Schliesser}},
  \bibinfo{author}{\bibfnamefont{P.}~\bibnamefont{Del'Haye}},
  \bibinfo{author}{\bibfnamefont{N.}~\bibnamefont{Nooshi}},
  \bibinfo{author}{\bibfnamefont{K.}~\bibnamefont{Vahala}}, \bibnamefont{and}
  \bibinfo{author}{\bibfnamefont{T.}~\bibnamefont{Kippenberg}},
  \bibinfo{journal}{Physical Review Letters} \textbf{\bibinfo{volume}{97}},
  \bibinfo{pages}{243905} (\bibinfo{year}{2006}).

\bibitem[{\citenamefont{Schliesser
  et~al.}(2008{\natexlab{b}})\citenamefont{Schliesser, Rivi{\`e}re,
  Anetsberger, Arcizet, and Kippenberg}}]{Schliesser2008}
\bibinfo{author}{\bibfnamefont{A.}~\bibnamefont{Schliesser}},
  \bibinfo{author}{\bibfnamefont{R.}~\bibnamefont{Rivi{\`e}re}},
  \bibinfo{author}{\bibfnamefont{G.}~\bibnamefont{Anetsberger}},
  \bibinfo{author}{\bibfnamefont{O.}~\bibnamefont{Arcizet}}, \bibnamefont{and}
  \bibinfo{author}{\bibfnamefont{T.}~\bibnamefont{Kippenberg}},
  \bibinfo{journal}{Nature Physics} \textbf{\bibinfo{volume}{4}},
  \bibinfo{pages}{415} (\bibinfo{year}{2008}{\natexlab{b}}).

\bibitem[{\citenamefont{Caves}(1981)}]{Caves1981}
\bibinfo{author}{\bibfnamefont{C.~M.} \bibnamefont{Caves}},
  \bibinfo{journal}{Physical Review D} \textbf{\bibinfo{volume}{23}},
  \bibinfo{pages}{1693} (\bibinfo{year}{1981}).

\bibitem[{\citenamefont{Kippenberg et~al.}(2005)\citenamefont{Kippenberg,
  Rokhsari, Carmon, Scherer, and Vahala}}]{Kippenberg2005}
\bibinfo{author}{\bibfnamefont{T.~J.} \bibnamefont{Kippenberg}},
  \bibinfo{author}{\bibfnamefont{H.}~\bibnamefont{Rokhsari}},
  \bibinfo{author}{\bibfnamefont{T.}~\bibnamefont{Carmon}},
  \bibinfo{author}{\bibfnamefont{A.}~\bibnamefont{Scherer}}, \bibnamefont{and}
  \bibinfo{author}{\bibfnamefont{K.~J.} \bibnamefont{Vahala}},
  \bibinfo{journal}{Physical Review Letters} \textbf{\bibinfo{volume}{95}},
  \bibinfo{pages}{033901} (\bibinfo{year}{2005}).

\bibitem[{\citenamefont{Pinard et~al.}(1999)\citenamefont{Pinard, Hadjar, and
  Heimann}}]{Pinard1999}
\bibinfo{author}{\bibfnamefont{M.}~\bibnamefont{Pinard}},
  \bibinfo{author}{\bibfnamefont{Y.}~\bibnamefont{Hadjar}}, \bibnamefont{and}
  \bibinfo{author}{\bibfnamefont{A.}~\bibnamefont{Heimann}},
  \bibinfo{journal}{European Physics Journal D} \textbf{\bibinfo{volume}{7}},
  \bibinfo{pages}{107} (\bibinfo{year}{1999}).

\bibitem[{\citenamefont{Anetsberger et~al.}(2008)\citenamefont{Anetsberger,
  Rivi{\`e}re, Schliesser, Arcizet, and Kippenberg}}]{Anetsberger2008}
\bibinfo{author}{\bibfnamefont{G.}~\bibnamefont{Anetsberger}},
  \bibinfo{author}{\bibfnamefont{R.}~\bibnamefont{Rivi{\`e}re}},
  \bibinfo{author}{\bibfnamefont{A.}~\bibnamefont{Schliesser}},
  \bibinfo{author}{\bibfnamefont{O.}~\bibnamefont{Arcizet}}, \bibnamefont{and}
  \bibinfo{author}{\bibfnamefont{T.~J.} \bibnamefont{Kippenberg}},
  \bibinfo{journal}{Nature Photonics} \textbf{\bibinfo{volume}{2}},
  \bibinfo{pages}{627} (\bibinfo{year}{2008}).

\bibitem[{\citenamefont{Arcizet et~al.}(2009)\citenamefont{Arcizet,
  Rivi{\`e}re, Schliesser, and Kippenberg}}]{Arcizet2009a}
\bibinfo{author}{\bibfnamefont{O.}~\bibnamefont{Arcizet}},
  \bibinfo{author}{\bibfnamefont{R.}~\bibnamefont{Rivi{\`e}re}},
  \bibinfo{author}{\bibfnamefont{A.}~\bibnamefont{Schliesser}},
  \bibnamefont{and} \bibinfo{author}{\bibfnamefont{T.~J.}
  \bibnamefont{Kippenberg}}, \bibinfo{journal}{arxiv:0901.1292}
  (\bibinfo{year}{2009}).

\bibitem[{\citenamefont{Pohl et~al.}(2002)\citenamefont{Pohl, Liu, and
  Thompson}}]{Pohl2002}
\bibinfo{author}{\bibfnamefont{R.~O.} \bibnamefont{Pohl}},
  \bibinfo{author}{\bibfnamefont{X.}~\bibnamefont{Liu}}, \bibnamefont{and}
  \bibinfo{author}{\bibfnamefont{E.}~\bibnamefont{Thompson}},
  \bibinfo{journal}{Review of Modern Physics} \textbf{\bibinfo{volume}{74}},
  \bibinfo{pages}{991} (\bibinfo{year}{2002}).

\bibitem[{\citenamefont{Braginsky and Vyatchanin}(2002)}]{Braginsky2002}
\bibinfo{author}{\bibfnamefont{V.~B.} \bibnamefont{Braginsky}}
  \bibnamefont{and} \bibinfo{author}{\bibfnamefont{S.~P.}
  \bibnamefont{Vyatchanin}}, \bibinfo{journal}{Physics Letters A}
  \textbf{\bibinfo{volume}{293}}, \bibinfo{pages}{228} (\bibinfo{year}{2002}).

\bibitem[{\citenamefont{Wineland and Itano}(1979)}]{Wineland1979}
\bibinfo{author}{\bibfnamefont{D.~J.} \bibnamefont{Wineland}} \bibnamefont{and}
  \bibinfo{author}{\bibfnamefont{W.~M.} \bibnamefont{Itano}},
  \bibinfo{journal}{Physical Review A} \textbf{\bibinfo{volume}{20}},
  \bibinfo{pages}{1521} (\bibinfo{year}{1979}).

\bibitem[{\citenamefont{Wilson-Rae et~al.}(2007)\citenamefont{Wilson-Rae,
  Nooshi, Zwerger, and Kippenberg}}]{Wilson-Rae2007}
\bibinfo{author}{\bibfnamefont{I.}~\bibnamefont{Wilson-Rae}},
  \bibinfo{author}{\bibfnamefont{N.}~\bibnamefont{Nooshi}},
  \bibinfo{author}{\bibfnamefont{W.}~\bibnamefont{Zwerger}}, \bibnamefont{and}
  \bibinfo{author}{\bibfnamefont{T.~J.} \bibnamefont{Kippenberg}},
  \bibinfo{journal}{Physical Review Letters} \textbf{\bibinfo{volume}{99}},
  \bibinfo{eid}{093901} (\bibinfo{year}{2007}).

\bibitem[{\citenamefont{Marquardt et~al.}(2007)\citenamefont{Marquardt, Chen,
  Clerk, and Girvin}}]{Marquardt2007}
\bibinfo{author}{\bibfnamefont{F.}~\bibnamefont{Marquardt}},
  \bibinfo{author}{\bibfnamefont{J.~P.} \bibnamefont{Chen}},
  \bibinfo{author}{\bibfnamefont{A.~A.} \bibnamefont{Clerk}}, \bibnamefont{and}
  \bibinfo{author}{\bibfnamefont{S.~M.} \bibnamefont{Girvin}},
  \bibinfo{journal}{Physical Review Letters} \textbf{\bibinfo{volume}{99}},
  \bibinfo{eid}{093902} (\bibinfo{year}{2007}).

\bibitem[{\citenamefont{Braginsky and Khalili}(1996)}]{Braginsky1996}
\bibinfo{author}{\bibfnamefont{V.~B.} \bibnamefont{Braginsky}}
  \bibnamefont{and} \bibinfo{author}{\bibfnamefont{F.~Y.}
  \bibnamefont{Khalili}}, \bibinfo{journal}{Reviews of Modern Physics}
  \textbf{\bibinfo{volume}{68}}, \bibinfo{pages}{1} (\bibinfo{year}{1996}).

\bibitem[{\citenamefont{Clerk et~al.}(2008{\natexlab{a}})\citenamefont{Clerk,
  Marquardt, and Jacobs}}]{Clerk2008a}
\bibinfo{author}{\bibfnamefont{A.~A.} \bibnamefont{Clerk}},
  \bibinfo{author}{\bibfnamefont{F.}~\bibnamefont{Marquardt}},
  \bibnamefont{and} \bibinfo{author}{\bibfnamefont{K.}~\bibnamefont{Jacobs}},
  \bibinfo{journal}{New Journal of Physics} \textbf{\bibinfo{volume}{10}},
  \bibinfo{pages}{095010} (\bibinfo{year}{2008}{\natexlab{a}}).

\bibitem[{\citenamefont{Genes et~al.}(2008)\citenamefont{Genes, Vitali,
  Tombesi, Gigan, and Aspelmeyer}}]{Genes2008}
\bibinfo{author}{\bibfnamefont{C.}~\bibnamefont{Genes}},
  \bibinfo{author}{\bibfnamefont{D.}~\bibnamefont{Vitali}},
  \bibinfo{author}{\bibfnamefont{P.}~\bibnamefont{Tombesi}},
  \bibinfo{author}{\bibfnamefont{S.}~\bibnamefont{Gigan}}, \bibnamefont{and}
  \bibinfo{author}{\bibfnamefont{M.}~\bibnamefont{Aspelmeyer}},
  \bibinfo{journal}{Physical Review A} \textbf{\bibinfo{volume}{77}},
  \bibinfo{pages}{033804} (\bibinfo{year}{2008}).

\bibitem[{\citenamefont{Clerk et~al.}(2008{\natexlab{b}})\citenamefont{Clerk,
  Devoret, Girvin, Marquardt, and Schoelkopf}}]{Clerk2008}
\bibinfo{author}{\bibfnamefont{A.~A.} \bibnamefont{Clerk}},
  \bibinfo{author}{\bibfnamefont{M.~H.} \bibnamefont{Devoret}},
  \bibinfo{author}{\bibfnamefont{S.~M.} \bibnamefont{Girvin}},
  \bibinfo{author}{\bibfnamefont{F.}~\bibnamefont{Marquardt}},
  \bibnamefont{and} \bibinfo{author}{\bibfnamefont{R.~J.}
  \bibnamefont{Schoelkopf}}, \bibinfo{journal}{arXiv:0810.4729}
  (\bibinfo{year}{2008}{\natexlab{b}}).

\bibitem[{\citenamefont{Ilchenko et~al.}(2004)\citenamefont{Ilchenko,
  Savchenkov, Matsko, and Maleki}}]{Ilchenko2004}
\bibinfo{author}{\bibfnamefont{V.~S.} \bibnamefont{Ilchenko}},
  \bibinfo{author}{\bibfnamefont{A.~A.} \bibnamefont{Savchenkov}},
  \bibinfo{author}{\bibfnamefont{A.~B.} \bibnamefont{Matsko}},
  \bibnamefont{and} \bibinfo{author}{\bibfnamefont{L.}~\bibnamefont{Maleki}},
  \bibinfo{journal}{Physical Review Letters} \textbf{\bibinfo{volume}{92}},
  \bibinfo{pages}{043903} (\bibinfo{year}{2004}).

\bibitem[{\citenamefont{Fabre et~al.}(1994)\citenamefont{Fabre, Pinard,
  Bourzeix, Heidmann, Giacobino, and Reynaud}}]{Fabre1994}
\bibinfo{author}{\bibfnamefont{C.}~\bibnamefont{Fabre}},
  \bibinfo{author}{\bibfnamefont{M.}~\bibnamefont{Pinard}},
  \bibinfo{author}{\bibfnamefont{S.}~\bibnamefont{Bourzeix}},
  \bibinfo{author}{\bibfnamefont{A.}~\bibnamefont{Heidmann}},
  \bibinfo{author}{\bibfnamefont{E.}~\bibnamefont{Giacobino}},
  \bibnamefont{and} \bibinfo{author}{\bibfnamefont{S.}~\bibnamefont{Reynaud}},
  \bibinfo{journal}{Physical Review A} \textbf{\bibinfo{volume}{49}},
  \bibinfo{pages}{1337} (\bibinfo{year}{1994}).

\bibitem[{\citenamefont{Mancini and Tombesi}(1994)}]{Mancini1994}
\bibinfo{author}{\bibfnamefont{S.}~\bibnamefont{Mancini}} \bibnamefont{and}
  \bibinfo{author}{\bibfnamefont{P.}~\bibnamefont{Tombesi}},
  \bibinfo{journal}{Physical Review A} \textbf{\bibinfo{volume}{49}},
  \bibinfo{pages}{4055} (\bibinfo{year}{1994}).

\bibitem[{\citenamefont{Vitali et~al.}(2007)\citenamefont{Vitali, Gigan,
  Ferreira, Bohm, Tombesi, Guerreiro, Vedral, Zeilinger, and
  Aspelmeyer}}]{Vitali2007}
\bibinfo{author}{\bibfnamefont{D.}~\bibnamefont{Vitali}},
  \bibinfo{author}{\bibfnamefont{S.}~\bibnamefont{Gigan}},
  \bibinfo{author}{\bibfnamefont{A.}~\bibnamefont{Ferreira}},
  \bibinfo{author}{\bibfnamefont{H.~R.} \bibnamefont{Bohm}},
  \bibinfo{author}{\bibfnamefont{P.}~\bibnamefont{Tombesi}},
  \bibinfo{author}{\bibfnamefont{A.}~\bibnamefont{Guerreiro}},
  \bibinfo{author}{\bibfnamefont{V.}~\bibnamefont{Vedral}},
  \bibinfo{author}{\bibfnamefont{A.}~\bibnamefont{Zeilinger}},
  \bibnamefont{and}
  \bibinfo{author}{\bibfnamefont{M.}~\bibnamefont{Aspelmeyer}},
  \bibinfo{journal}{Physical Review Letters} \textbf{\bibinfo{volume}{98}},
  \bibinfo{pages}{030405} (\bibinfo{year}{2007}).

\bibitem[{\citenamefont{Heidmann et~al.}(1997)\citenamefont{Heidmann, Hadjar,
  and Pinard}}]{Heidmann1997}
\bibinfo{author}{\bibfnamefont{A.}~\bibnamefont{Heidmann}},
  \bibinfo{author}{\bibfnamefont{Y.}~\bibnamefont{Hadjar}}, \bibnamefont{and}
  \bibinfo{author}{\bibfnamefont{M.}~\bibnamefont{Pinard}},
  \bibinfo{journal}{Applied Physics B} \textbf{\bibinfo{volume}{64}},
  \bibinfo{pages}{173} (\bibinfo{year}{1997}).

\bibitem[{\citenamefont{Kippenberg and Vahala}(2007)}]{Kippenberg2007}
\bibinfo{author}{\bibfnamefont{T.~J.} \bibnamefont{Kippenberg}}
  \bibnamefont{and} \bibinfo{author}{\bibfnamefont{K.}~\bibnamefont{Vahala}},
  \bibinfo{journal}{Optics Express} \textbf{\bibinfo{volume}{15}},
  \bibinfo{pages}{17172} (\bibinfo{year}{2007}).

\end{thebibliography}

\end{document}